\documentclass[%
 reprint,
 amsmath,amssymb,
 aps,
]{revtex4-1}

\usepackage{graphicx}
\usepackage{dcolumn}
\usepackage{bm}
\usepackage[T1]{fontenc}
\usepackage{multirow}
\usepackage{url}

\begin{document}


\title{ Modeling and analysis of the effect of COVID-19 on the stock price: V and L-shape recovery}

\author{Ajit Mahata, Anish Rai and Md Nurujjaman}
\email{md.nurujjaman@nitsikkim.ac.in}
 \affiliation{Department of Physics, National Institute of Technology Sikkim, Sikkim, India-737139.}
\author{Om Prakash}
\affiliation{Department of Mathematics, National Institute of Technology Sikkim, Sikkim, India-737139.}%

\date{\today}

\begin{abstract}
The emergence of the COVID-19 pandemic, a new and novel risk factor, leads to the stock price crash due to the investors' rapid and synchronous sell-off. However, within a short period, the quality sectors start recovering from the bottom. A stock price model has been developed to capture the price dynamics during shock and recovery phases of such crisis. The main variable and parameter of the model are the net-fund-flow ($\Psi_t$) due to institutional investors, and financial antifragility ($\phi$) of a company, respectively. We assume that during the crash, the stock price fall is independent of the $\phi$. We study the effects of shock length ($\mathcal{T_S}$) and $\phi$ on the stock price during the crisis period using the $\Psi_t$ obtained from both the synthetic fund flow data and real fund flow data. We observed that the possibility of recovery of stock with $\phi>0$, termed as quality stock, decreases with an increase in $\mathcal{T_S}$ beyond a specific period. A quality stock with higher $\phi$ shows V-shape recovery and outperform others. The $\mathcal{T_S}$ and recovery period of quality stock are almost equal in the Indian market. Financially stressed stocks, i.e., the stocks with $\phi<0$, show L-shape recovery during the pandemic. The stock data and model analysis show that the investors, in the uncertainty like COVID-19, invest in the quality stocks to restructure their portfolio to reduce the risk. The study may help the investors to make the right investment decision during a crisis.

\end{abstract}

\pacs{Valid PACS appear here}
\maketitle


\section{Introduction}
The impact of a pandemic on the environment, economy, employment, stock market and many other sectors is severe~\cite{muhammad2020covid,barro2020coronavirus,ashraf2020economic,topcu2020impact}. A number of pandemics, which happened in $1918-1919$, $1957-1958$ and 1968, affected the economy and stock markets worldwide badly~\cite{velde2020happened,jorda2020longer,baker2020unprecedented}. The pandemic's effect, the coronavirus disease (COVID-19), maybe more severe on the economy and stock market, including many other sectors, due to its contagion nature~\cite{baker2020unprecedented,goodell2020covid,narayan2020covid}. The COVID-19 leads to a worldwide stock market crash in February and March 2020, created havoc among the investors~\cite{mazur2020covid,zhang2020financial,bbc}. When economic activities throughout the world were plummeting, surprisingly, in the stock market, the world witnessed the opposite phenomena; the speedy recovery of some stocks and sectors from the crash~\cite{mazur2020covid}. These phenomena were also observed during market crashes in $1953-54$, 2009~\cite{forbes}. The study of market crash dynamics and early recovery of the stocks during the crisis is fascinating.

The stocks with strong fundamentals and positive outlook, which are termed as quality stocks, always show strength, universality and persistence in returns~\cite{asness2019quality,bouchaud2016excess,novy2013other}. The sustainability and resilience of a quality stock depend on its long-term growth prospect and financial ability to fulfill shareholder's demand~\cite{castro2006integrated,lee2009corporate}. During crises, the unprecedented economic uncertainty forced the investors towards the quality stocks that may be better able to withstand a downturn, and hence the significant portion of the market capital gets reallocated to these sectors, which in turn pushes the price up~\cite{bloomberg_lyz}. Thus the sectors like pharma, healthcare, food, software and technology showed the quality of withstanding the downturn and recovered quickly from the sharp fall. Whereas sectors like petroleum, real estate, entertainment, hospitality yet to recover because of grim business outlook~\cite{mazur2020covid}.

The quality of a company can be quantified by the fundamental determinants such as profit over the assets, return on assets, operating cash-flows to total assets, gross margin, sale growth and some other fundamental determinants that assess the reliabilities of profits, low debt and other measures of sustainable earnings~\cite{asness2019quality,bouchaud2016excess,novy2013other}. Investors look for such quality stocks even at a high premium in anticipation of higher returns~\cite{asness2019quality}. The survival and growth of these quality stocks during pandemic depend on the financial antifragility of the company. Financial antifragility is the property that shows the ability of a company to survive from a financial crisis and performs strongly after that, and it mainly depends on the financial liquidity position to mitigate the liabilities~\cite{taleb2013mathematical,taleb2012antifragile}. The stocks with positive antifragility recover very quickly from uncertain shock and survive and sustain for an extended period~\cite{taleb2013mathematical,taleb2012antifragile,platje2015sustainability}. 

The fund-flow in the market is primarily determined by the foreign institutional investors (FII) and domestic institutional investors (DII). The purchase/sell activity by the FII and DII also influence the retail investors~\cite{cao2008empirical}. Hence, the net fund-flow by the FII and DII drives the stock price~\cite{coval2007asset,ulku2014identifying,kling2008chinese,edelen2001aggregate,KP2020Dynamics}. Infusion of a large amount of fund to a particular sector leads to an increase in the stock price or vice versa, i.e., the price movement is strongly correlated with the net fund flow due to FII and DII~\cite{KP2020Dynamics,edelen2001aggregate,edelen1999investor,warther1995aggregate}. Generally, during a crisis, the FII and DII look for stocks with robust financial antifragility, and hence these stocks bounce back strongly~\cite{bloomberg_lyz}. Hence modeling and analyzing stock prices in terms of net fund flow and antifragility during shock and recovery phases of a pandemic are essential to understanding the market dynamics.

Several models describe the stock price movement using various parameters such as return and dividend ~\cite{ding1993long,marsh1987dividend,marsh1986dividend}. The efficient market hypothesis (EMH) states that the future price does not depend on the past behavior of data~\cite{fama1970efficient}. Contrary to the EMH hypothesis, some other models show that the stock price is partially predictable~\cite{malkiel2003efficient}. However, the stock price prediction remains a challenging task due to the stock market's complex nature ~\cite{ballings2015evaluating}. Recently, a model of V and L shape recovery of the economy is proposed in Ref~\cite{sharma2020v,gualdi2015tipping} depending on the fragility of the individual firms, where the fragility was taken as a ration between negative cash balance to the wages. So far, no one has modeled and analyzed stock prices in terms of antifragility and fund flow. 

The main aim of this paper is to develop a model, and to simulate stock price movement during the COVID-19 shock and subsequent recovery as a function of normalized net fund flow ($\Psi_t$) due to institutional investors and the antifragility parameter of a stock. Model simulation has been carried out for two different sets of $\Psi_t$: (a) $\Psi_t$ obtained from the real fund flow in the market, and (b) artificially generated $\Psi_t$ using the distribution of net cash flow. Simulation with real fund flow reproduces the price movement of the quality stocks and financially stressed stocks during the COVID-19 shock that mimics the actual stock price.  The model simulation with artificial data shows the effect of various shock-lengths and antifragility parameters. Further, we have analyzed the stock price using EMD based Hilbert Huang transformation in terms of the time scales of the shock and recovery to identify the quality stocks~\cite{menkhoff2010use,mahata2020identification}. 

 The rest of the paper is organized as follows: Sec.~\ref{sec:model} describes the formulation of the model. Sec.~\ref{sec:rd} discusses the analysis of the simulated results and original stock price. Finally, we have concluded the results in Sec.~\ref{sec:con}.

\section{Model formulation}
\label{sec:model}
A model has been developed for the stock price dynamics of a stock/sector index during the shock and recovery period. The time steps of the model are discrete with a step of one day. The model's basic assumptions are that the stock price depends on (a) the net fund flow due to FII and DII, and (b) financial antifragility of the company. The basis of the first assumptions is motivated by the finding of the daily stock return is positively correlated with the net fund flow~\cite{edelen2001aggregate,edelen1999investor,warther1995aggregate}. The retail investors also flock towards the sector in which FII and DII invest more. Sometimes the price goes up or down with lag to net fund flow because of information delay~\cite{edelen2001aggregate,sun2016stock}. Hence, the overall market moves with net fund flow due to institutional investors. 

During shock, the market falls due to negative sentiment among the investors, leading to a huge outflow of capital from the market. As the pessimism dies down, the investors again come to the market, and invest in those companies which have strong fundamentals and positive growth prospects. Hence, the fund inflow happens to the company with positive antifragility~\cite{glossner2020institutional,fahlenbrach2020valuable}. The model with the variables normalized-net-fund-flow and antifragility can capture price dynamics very well during shock and recovery phase.

Let us first define the main ingredients of the model. The variable used in the model is the net fund flow, which can be calculated as follows. The cash purchase (inflow) or sell (outflow) by the $FII$ is denoted as $D_{FII}^+$ or $D_{FII}^-$, respectively. Hence, the net cash purchase by the $FII$ is $D_{FII}=D_{FII}^+-D_{FII}^-.$ Similarly, the cash purchase (inflow) or sell (outflow) by the $DII$ is denoted as $D_{DII}^+$ or $D_{DII}^-$, respectively. Hence, the net cash purchase by the $DII$ is $D_{DII}=D_{DII}^+-D_{DII}^-.$ So (the net fund flow due to DII) the net cash purchase is defined as $D_{DII}=D_{DII}^+-D_{DII}^-.$ Finally, we obtained $\Delta D_t$ due to the purchase or sale by the institutional investors is $\Delta D_t=D_{FII}+D_{DII}.$ Finally, we obtained normalized-net-fund-flow, 
\begin{equation}
\Psi_t= \frac{\Delta D_t}{max(abs(\Delta D_t))}
\label{eqn:psi}
\end{equation}
$\Psi_t$ is the variable that is used to update the stock price.

The second ingredient of the model is the antifragility parameter ($\phi$), which is estimated as follows: The recovery of a company after a shock mainly depends on current asset consumed to fulfill the current liabilities of the company. The asset that is used, sold, consumed or exhausted during a normal operating cycle is called the current assets of a company. The current assets can easily cover day-to-day financial operations and ongoing operating expenses; hence, it becomes a key component for a company's survival or death. The current assets of the $i^{th}$ company, $\chi_{ i}$, is defined as
$\chi_{ i}=\eta_{1i}+\eta_{2i}+\eta_{3i}+\eta_{4i},$
where $\eta_{1i}$, $\eta_{2i}$,  $\eta_{3i}$ and $\eta_{4i}$ are the current inventories, trade receivables, cash and cash equivalents and other current assets, respectively. Current liabilities are the obligations of a company that consists of short term debt and other similar debts that will be due within a normal operating cycle. Therefore, we define current liabilities of the $i^{th}$ company as  
$\zeta _{i}=\gamma_{1i}+\gamma_{2i}+\gamma_{3i},$ where $\gamma_{1i}$, $\gamma_{2i}$ and $\gamma_{3i}$ are the current debt, trade payable and other current liabilities respectively. Hence, the liquidity balance of the company is defined as $\chi_{i}-\zeta_i.$ We characterize the financial antifragility ($\phi$) of the $i^{th}$ company through liquidity-to-expense ratio

\begin{equation}
\displaystyle\phi_{i}=\frac{\chi_{i}-\zeta_{i}}{\xi_{i}}
\label{eqn:phi1}
\end{equation}
Where $\xi_{i}$ is the operating expenses of a company, and is expressed as $\xi_{i}=\vartheta_{1i}+\vartheta_{2i}+\vartheta_{3i}+\vartheta_{4i},$  where $\vartheta_{1i}$, $\vartheta_{2i}$, $\vartheta_{3i}$, and $\vartheta_{4i}$ are the employment cost, financial cost, maintenance and operating cost and other financial cost respectively. The $\phi$ for a sector can be written as 
\begin{equation}
\displaystyle\phi=\frac{\sum_{1}^{N} \phi_{i}}{N}
\label{eqn:phi2}
\end{equation}

Where N is number of company in any sector's index. $\phi$ acts as the control parameter of the price movement. The value of $\phi>0$ for a quality stock, and $\phi<0$ for a financially stressed stock. Usually, $\phi$ get updated twice a year based on the financial statement of a company. It is important to mention that during shock, market nose dives due to massive sell-off by the investors, and hence the price movement is independent of $\phi.$

Our model updates the stock price as a function of $\Psi_t$ using the parameter $\phi$ as follows
 
\begin{eqnarray}
P_{t+1}&=&P_t\{1+\lambda\Psi_t)\},~~~~~~During~shock\label{eqn:model1}\\
P_{t+1}&=&P_t\{1+\lambda\Psi_t\phi\},~~~~~~Otherwise
\label{eqn:model2}
\end{eqnarray} 			

Where, $\lambda$ is the coefficient of $\Psi_t$ that represents the proportion of the net fund by the institutional investors that flows in a particular company/sector. The value of $\lambda$ changes during normal, shock and recovery period. The value of $\lambda$ has been taken on adhoc basis depending on the normalized fund flow due to the mutual fund and FPI. Typically the value of $\lambda$ is in the range of $0 \leq\lambda \leq1.$

There are large numbers of companies and indices in the stock market. To understand the price movement of these companies and indices during the COVID-19 shock, one needs to study the model with different shock and recovery lengths and antifragility parameters. Hence, in Subsec.~\ref{subsec:covid}, the model equations [Eqn~\ref{eqn:model1} and \ref{eqn:model2}] is simulated using artificially generated fund flow data to understand the COVID-like shock. The artificial data is generated from the normal distribution of real fund flow during different phases of the COVID crisis. Further, the model is studied in Subsec.~\ref{subsec:covid-19} for the COVID-19 shock by using real fund flow and antifragility of the company.

\subsection{COVID-like shock}
\label{subsec:covid}
Study of the effect of the COVID-like shock on the stock prices in terms of various shock lengths ($\mathcal{T_S}$) and different $\phi$ is very important to understand the market crash and subsequent recovery. We have generated synthetic normalized fund flow ($\Psi_{st}$) data from the distribution of $\Psi_t$ [Eqn.~\ref{eqn:psi}] during the normal, shock and recovery period.
The distribution of $\Psi_t$ for the normal, shock and recovery periods are $\mathcal{N}(0,0.17)$, $\mathcal{N}(-0.2,0.49)$ and $\mathcal{N}(0.06,0.22)$, respectively, and accordingly $\Psi_{st}$ is generated. As the distribution of the $\Psi_t$ is derived from real data, the $\Psi_{st}$ mimics the real situation. In this model simulation, the recovery time period ($\mathcal{T_R}$) is taken equal to $\mathcal{T_S}$, and the justification is given in detail in Subsec.~\ref{subsec:timescale}. The value of $\lambda=0.2,~0.1,~0.7,~0.3$ during pre-covid normal period, shock period, recovery period and post recovery period, respectively. The reason for choosing different values of $\lambda$ during different periods is discussed in Sec.\ref{subsec:covid-19}.

To understand the V-shape recovery of a quality stock, the model simulation is carried out for the fixed $\phi=0.4$ with $\mathcal{T_S}=20~day~(D),~40~D,~60~D,$ and $80~D$, respectively, and  for the fixed $\mathcal{T_S}=20~D$ with $\phi=0.3,~0.4,~0.5$ and $0.6$. The value of $\phi$ and $\mathcal{T_S}$ are chosen based on the original stock price. Similarly, for the L-shape recovery of the financially stressed company, simulation is carried for the $\phi=-0.08$ with $\mathcal{T_S}=20~D,~40~D,~60~D,$ and $80~D$, respectively, and for the  $\mathcal{T_S}=20~D$ with $\phi=-0.05,~-0.15,~-0.25$ and $-0.35.$ In this case, the value of $\phi$ and $\mathcal{T_S}$ are chosen on the basis of the original stock price. The detailed simulation result is given in Sec.~\ref{sec:rd}.

\subsection{COVID-19 shock}
\label{subsec:covid-19}

The model simulation has been carried out for the stock price during the COVID-19 using the $\Psi_t$ and $\phi$ for the Indian market. The $\Psi_t$ has been calculated from real net fund flow in the market due to FII and DII using Eqn.~\ref{eqn:psi}. The fund flow data has been obtained from the money control website~\cite{moneycontrol}. The current financial status of a company, $\phi$, is estimated using Eqn.~\ref{eqn:phi1}. The value of $\phi$ for a sector has been calculated using Eqn.~\ref{eqn:phi2}. The current assets, current liabilities and expenses of a company have been derived from its financial statements, which are obtained from Bombay Stock Exchange Ltd (BSE)~\cite{bse}.

\begin{figure}
\includegraphics[angle=0, width=7.4cm]{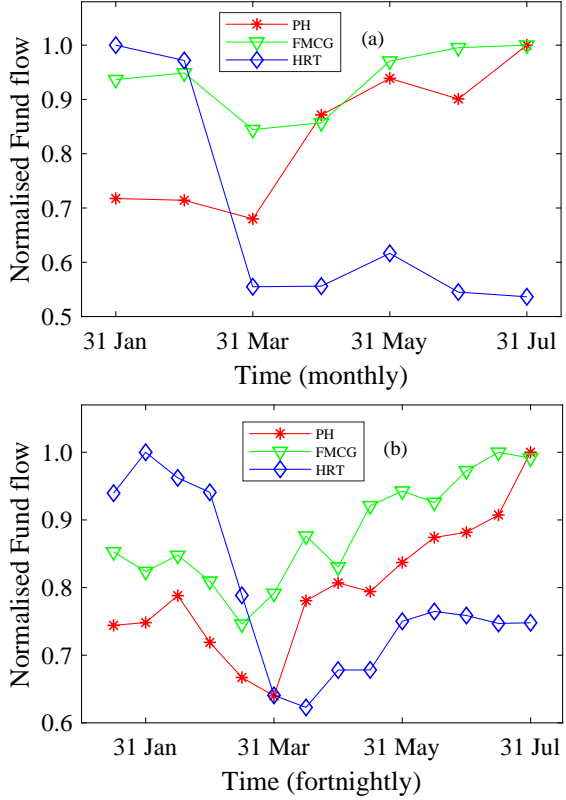}
\caption{\label{fig:psi}Plot (a) shows the monthly data of the normalized fund flow due to mutual fund in Pharma ($-*-PH$), FMCG ($-\nabla-FMCG$) and Hotel and Tourism ($-\diamond-HRT$) sectors. Plot (b) shows the fortnightly data of the normalized fund flow due to FPI in Pharma ($-*-PH$), FMCG ($-\nabla-FMCG$) and Hotel and Tourism ($-\diamond-HRT$) sectors.}
\end{figure}

The coefficient $\lambda$ for a particular stock depends on the ratio of fund flow to total fund flow in the market. In the present case, the value of $\lambda$ is taken in adhoc manner that can be guessed from the fund flow to a sector due to the institutional investors. Fig.~\ref{fig:psi}(a)  shows the normalized monthly fund flow due to mutual fund investors (MFI) in Pharma \& Biotechnology ($-*-$PH), Fast Moving Consumers good($-\nabla-$FMCG) and Hotel Restaurant and tourism($-\diamond-$HRT). The above sector data has been obtained from the Securities and Exchange Board of India (SEBI)~\cite{sebi}. Fig.~\ref{fig:psi}(b) shows the normalized fortnightly fund flow due to foreign portfolio investors (FPI) in Pharma \& Biotechnology ($-*-$PH), Fast Moving Consumers good($-\nabla-$FMCG) and Hotel Restaurant and tourism($-\diamond-$HRT), and the data has been obtained from National Securities Depository Ltd., India (NSDL)~\cite{nsdl}. The figure shows that the fund flow during shock decreased significantly, and during the recovery phase, fund flow in the pharma and FMCG sectors increased significantly. On the other hand, in the Hotel Restaurant and tourism sector, fund flow remains almost constant after a drastic drop. Considering the above information, for the quality stock, we have taken $\lambda=0.6,~0.2,~0.8,~0.6$ for Nifty Pharma and $\lambda=0.6,~04,~0.9,~0.7$ for Nifty FMCG index during the pre-COVID normal period, shock period, recovery period and post-recovery period, respectively, and for a financially stressed stock, like Tata Motors, $\lambda=0.6,~1.0,~0.8,~0.7$ and for BPCL $\lambda=0.6,~0.8,~0.8,~0.7$ during the above four periods. The detailed simulation result is given in Sec.~\ref{sec:rd}.

\section{Results and discussion}
\label{sec:rd}
This section aims to present the simulation results of the effect of the $\mathcal{T_S}$ and $\phi$ on the V- and L-shape recovery of the stock price. The analysis of the original price and the simulated price has also been carried out to identify the shock and recovery time scales of the stock price.

\subsection{Simulation of COVID-like shock}

\begin{figure}
\includegraphics[angle=0, width=7.2cm]{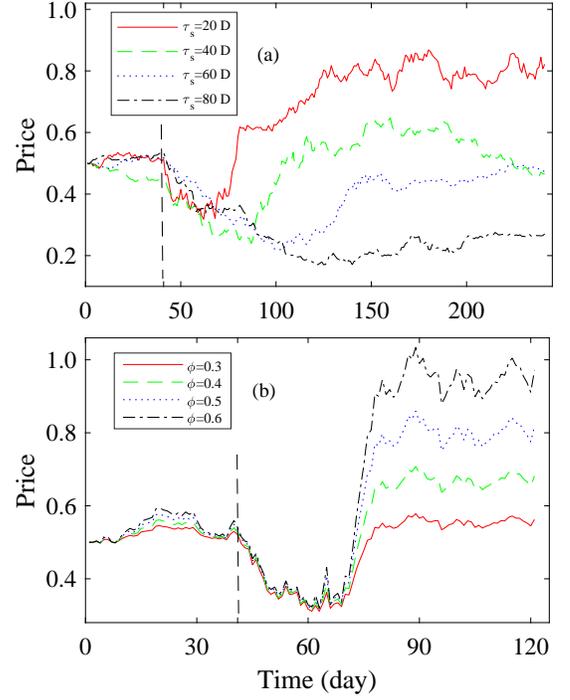}
\caption{\label{fig:SD} Plot (a) shows the V-shape recovery using the synthetic data for fixed $\phi=0.4$ with $\mathcal{T_S}=20~D,~40~D,~60~D$ and $80~D$. Plot (b) shows the V-shape recovery using synthetic data for fixed $\mathcal{T_S}=20~D$ with $\phi=0.3,~0.4,~0.5$ and 0.6. $D$ and $\phi$ represent day and financial antifragility. Vertical dashed line represents the starting point of shock. For the simulation initial condition is taken as 0.5}
\end{figure}

Fig.~\ref{fig:SD}(a) shows the typical plot of V-shape recovery of a quality stock using $\Psi_{st}$ for different $\mathcal{T_S}$ with $\phi=0.4.$ As the typical value of $\phi$ for a quality sector is around 0.4. The plot $-$, $--$, $\cdots$, and $-\cdot-$ show the stock price movement for the $\mathcal{T_S}=20~D,~40~D,~60~D,$ and $80~D$, respectively. The results show that the quality stock recovers very well to its pre-shock price for the $\mathcal{T_S}=~20~D$ and $40~D$. However, when the shock extended beyond $60~D$, it becomes difficult to recover because of the stock price's serious crash. It implies that the extended period of $\mathcal{T_S}$ is harmful even for the financially strong company. During such kind of extended shock, the investor stays away from investment in the market, which is sometimes termed by the investors "Do not catch a falling knife"~\cite{colliard2013catching}.

Fig.~\ref{fig:SD}(b) shows the typical plot of V-shape recovery of quality stocks for different $\phi$ with $\mathcal{T_S}=20~D.$ The typical value of $\mathcal{T_S}$ was 20 D during the COVID-19 for a quality sector. The plot $-$, $--$, $\cdots$, and $-\cdot-$ show the stock price movement for the $\phi=0.3,~0.4,~0.5$ and $0.6$, respectively. The higher the value of $\phi$, the recovery is rapid. The results show that the quality stock recovers very well for positive $\phi$. Further, the quality stocks with higher $\phi$ outperform its peer. Hence, the financially strong company recover from the shock and outperform compared to another company observed during the COVID-19~\cite{glossner2020institutional,fahlenbrach2020valuable}. During crises, the investors invest heavily in such companies, and hence generates higher return~\cite{bloomberg}.

\begin{figure}
\includegraphics[width=7.5cm]{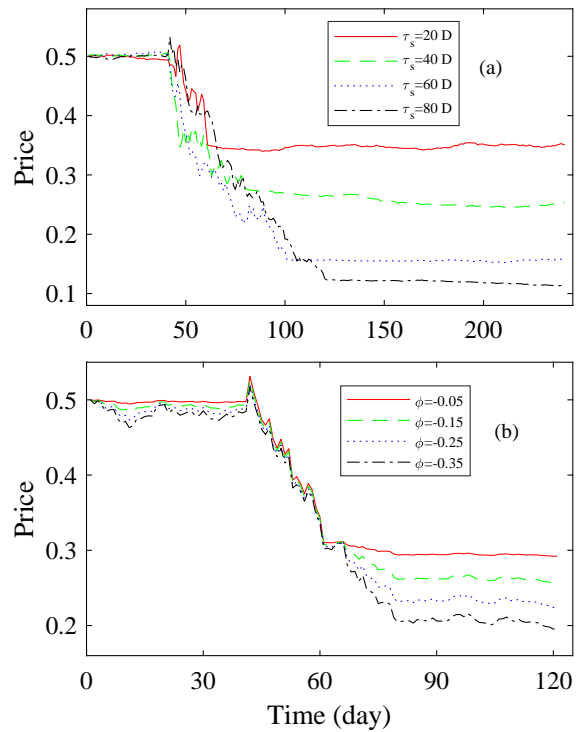}
\caption{\label{fig:OD_Simu} Plot (a) shows the L-shape recovery using the synthetic data for fixed $\phi=-0.08$ with $\mathcal{T_S}=20~D,~40~D,~60~D$ and $80~D$. Plot (b) shows the L-shape recovery using the synthetic data for fixed $\mathcal{T_S}=20~D$ with $\phi=-0.05,~-0.15,~-0.25$ and -0.35. For the simulation initial condition is taken as 0.5. $D$ and $\phi$ represent day and financial antifragility.}
\end{figure}

Fig.~\ref{fig:OD_Simu}(a) shows the typical plot of L-shape recovery of a financially stressed stock for different $\mathcal{T_S}$ with $\phi=-0.08.$ Typical value of $\phi$ for this sector is around $-0.08$. The plot $-$, $--$, $\cdots$, and $-\cdot-$ show the stock price movement for the $\mathcal{T_S}=20~D,~40~D,~60~D,$ and $80~D$ respectively. The simulation results show that the financially stressed stock does not recover. As the value of $\mathcal{T_S}$ increases, the negative depth of stock price also increases. So, a financially stressed company cannot survive the extended $\mathcal{T_S}$, and have a big chance to die down. The investors become very bearish on these company, and sell-off their positions, and hence the chance of the recovery of the stock price also becomes marginal.

Fig.~\ref{fig:OD_Simu}(b) also shows the typical plot of the L-shape behavior of a financially fragile stock for different $\phi$ with $\mathcal{T_S}=20~D.$  The plot $-$, $--$, $\cdots$, and $-\cdot-$ show the stock price movement for the $\phi=-0.05,~-0.15,~-0.25$ and $-0.35$, respectively. For the simulation of COVID-like shock we have taken 0.5 as initial condition. The company with negative $\phi$ continues to slide down even during the recovery phase of the overall market. The lower the $\phi$, the slide in stock price is rapid. So, a company with a lower $\phi$ has a big chance to die down. As the investors stay away from these companies, the chance of the stock price recovery also becomes marginal as mentioned in the previous paragraph.

\subsection{COVID-19 shock}

\begin{figure}
\includegraphics[angle=0, width=7.5cm]{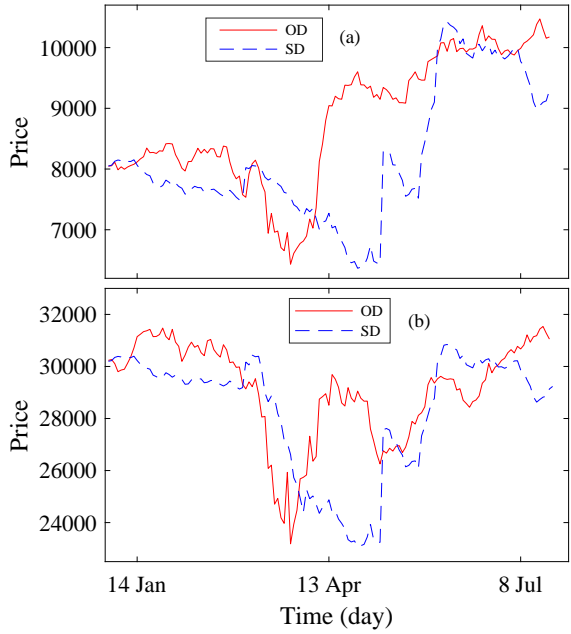}
\caption{\label{fig:OD_Simu1} Plot (a) represents the original stock price movement of Nifty Pharma ($-OD$) and its corresponding model simulated stock price movement ($--SD$) with $\phi=0.41 $ and $\mathcal{T_S}=20~D$. Plot (b) represents the original stock price movement of Nifty  FMCG ($-OD$) and its corresponding model simulated stock price movement ($--SD$) with $\phi=0.21$ and $\mathcal{T_S}=20~D$}
\end{figure}

Fig.~\ref{fig:OD_Simu1}(a) and \ref{fig:OD_Simu1}(b) show the simulation result ($--$) of the model using the  normalized fund flow $\Psi_t$ given in Eqn~\ref{eqn:psi}, and $-$ represents stock price of the Pharma and FMCG index in Indian market during the COVID-19 shock. The simulation results of Pharma and FMCG indices show that the fall of the stock price due to the COVID-19 shock starts from $1^{st}$ week of March 2020, and forms a bottom on $4^{th}$ week of April 2020, as shown in Fig.~\ref{fig:OD_Simu1}(a) and~\ref{fig:OD_Simu1}(b), respectively. The model simulation shows a V-shape recovery in the stock price consistent with the original stock price during the shock, as shown in the same figure. We observed a lag in the formation of the bottom between the model simulation data (SD) and original data (OD). Original stock price recovers earlier than the model. The possible reason for such lag may be due to the fund allocation in the quality sectors by the investors internally, which was not reflected in the fund flow. For example, in India, during the pandemic, the outlook in the Pharma and FMCG sectors becomes positive, hence the fund allocation to these sectors due to DII and retail investors increased rapidly that can be understood only from investors' buying sentiment. The model for the quality index and company may behave properly with minimum lag if index or company wise fund flow data were available. 

\begin{figure}
\includegraphics[angle=0, width=7.5cm]{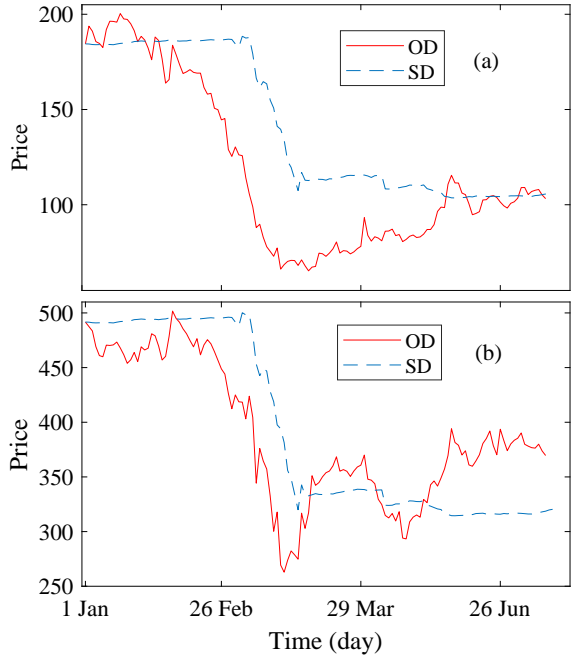}
\caption{\label{fig:lshape} Plot (a) shows the original stock price movement of Tata Motors Ltd ($-OD$) and its corresponding model simulated stock price movement ($--SD$) with $\phi=-0.077 $ and $\mathcal{T_S}=20~D$. Similarly, plot (b) shows the original stock price movement of BPCL ($-OD$) and its corresponding model simulated stock price movement ($--SD$) with $\phi=-0.052$ and $\mathcal{T_S}=20~D$.}
\end{figure}

\begin{figure}
\includegraphics[angle=0, width=7.5cm]{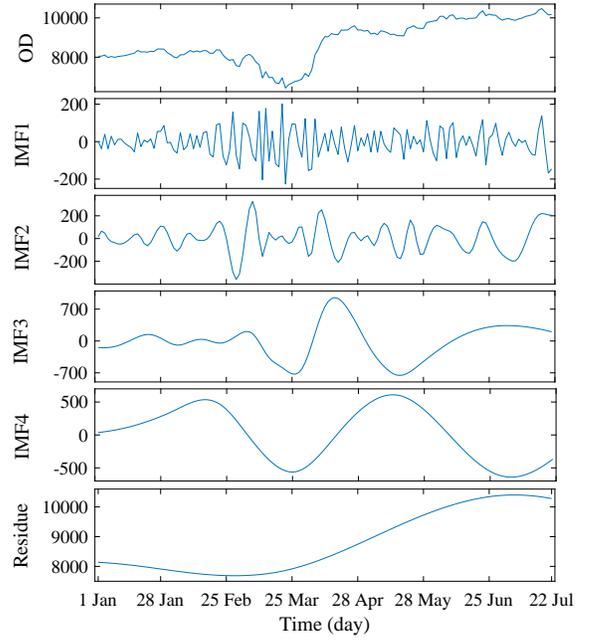}
\caption{\label{fig:IMF} Original data (OD) of Nifty Pharma and its IMF and residue obtained using empirical mode decomposition (EMD) technique.}
\end{figure}
  
Fig.~\ref{fig:lshape}(a) and \ref{fig:lshape}(b) show the simulation result for the stock price that show L-shape recovery in Indian market during COVID-19 shock. The simulation for Tata Motors and BPCL show that the fall of the stock price due to the COVID-19 shock starts from $1^{st}$ week of March 2020, and forms a bottom on $4^{th}$ week of March 2020 as shown in Fig.~\ref{fig:lshape}(a) and~\ref{fig:lshape}(b), respectively. We observed that the stock with $\phi<0$ does not recover from the bottom, i.e., the behavior of the price is L-shape, as shown in the same figure. The main reasons for the poor allocation of the fund, as shown in Fig.~\ref{fig:psi} in these stressed stocks, are the non-essential nature of the product they produce, negative outlook during COVID-19.  

\subsection{Time scale Analysis}
\label{subsec:timescale}

The empirical mode decomposition (EMD) technique is applied to identify the important $\mathcal{T_S}$ and $\mathcal{T_R}$ of stocks and indices during the COVID-19~\cite{huang1998empirical,huang2014hilbert}. The EMD technique decomposes a signal into a number of intrinsic mode functions (IMF) of different time scales by preserving the nonlinearity and nonstationarity of a time series~\cite{huang2003applications}. The detailed algorithms for identifying the IMF using the EMD method is given in Ref.~\cite{huang1998empirical,mahata2020identification}.

The range of a time period of a particular IMF can be obtained using $\displaystyle\tau=\frac{1}{\omega}$, where $\displaystyle\omega=\frac{d\theta(t)}{dt}$. The $\omega$ of a particular  $IMF$ can be estimated by using Hilbert Transform, which is defined as 
$$\displaystyle Y(t)=\frac{P}{\pi}\int_{-\infty}^{\infty}\frac{IMF(t)}{t-t'}dt,$$ where $P$ is the Cauchy principle value, and $\displaystyle \theta(t)=tan^{-1}\frac{Y(t)}{IMF(t)}$~\cite{huang1998empirical}. We have applied EMD based Hilbert Huang Transformation to obtain the $\tau$ of the stock data.

We have identified the $\tau$ of the Nifty Pharma and Nifty FMCG index as quality stocks, and their model simulated data during the COVID-19 shock. The data were taken from~\cite{yahoo,nse}.

Fig.~\ref{fig:IMF} shows the IMF of the Nifty Pharma index estimated using the EMD technique. The series is decomposed into four IMFs and a residue. Fig.~\ref{fig:IMF} shows the visualization. IMF1 represents the signal with the lowest $\tau$, and the $\tau$ increases with the increase in IMF numbers. The residue represents the overall long-term trend of the index. Each IMF represents a mono-frequency component of the stock data. 

\begin{table*}
\caption{Measures of correlation coefficient ($\nu$) between original data (OD) and its IMFs and model simulated data (SD) and its IMFs of NIfty Pharma and Nifty FMCG index. Variance ($\sigma^2$) has been calculated for the IMFs of OD and SD.}
\label{tab:1}
\begin{tabular}{|c|c|c|c|c|c|c|c|c|}
\hline
\multirow{2}{*}{IMF NO.} & \multicolumn{4}{c|}{ Nifty Pharma index} & \multicolumn{4}{c|}{Nifty FMCG index}  \\
\cline{2-9}
 & \multicolumn{2}{c|}{OD} & \multicolumn{2}{c|}{SD} & \multicolumn{2}{c|}{OD} & \multicolumn{2}{c|}{SD} \\
\cline{2-9}
 & $\nu$ & $\sigma^2$ & $\nu$ & $\sigma^2$ & $\nu$ & $\sigma^2$ & $\nu$ & $\sigma^2$  \\
\hline
 IMF1&0.1456&8.81$\times10^{3}$&0.0605&2.64$\times10^{4}$&0.2681&$1.62\times10^{5}$&0.0139&1.32$\times10^{5}$ \\
\hline
 IMF2&0.0576&1.75$\times10^{4}$&0.2997&$1.05\times10^{5}$&0.1394&7.65$\times10^{4}$&0.2238&3.99$\times10^{5}$ \\
\hline
 IMF3&0.4219&2.44$\times10^{5}$&0.3504&2.29$\times10^{4}$&0.5159&1.68$\times10^{6}$&0.2681&1.88$\times10^{5}$ \\
\hline
 IMF4&0.7394&1.76$\times10^{5}$&0.8544&2.94$\times10^{5}$&0.7475&1.33$\times10^{6}$&0.8724&1.69$\times10^{6}$ \\
\hline
\end{tabular}
\end{table*}
	
\begin{figure*}
\includegraphics[angle=0, width=15cm]{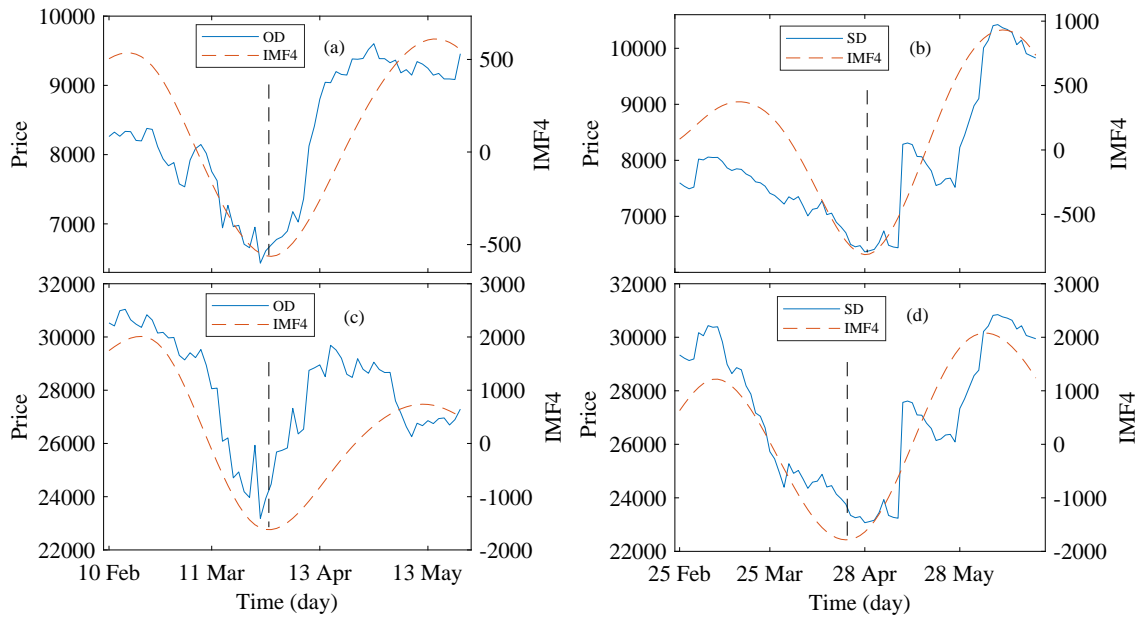}
\caption{\label{fig:tscaleV} Plot (a) shows the original data (OD) of Nifty Pharma and its dominant IMF (IMF4). Plot (b) shows the model simulated data (SD) of Nifty Pharma and its dominant IMF (IMF4). Plot (c) shows the original data (OD) of Nifty  FMCG and its dominant IMF (IMF4). Similarly, plot (d) shows the model simulated data (SD) of Nifty  FMCG and its dominant IMF (IMF4).}
\end{figure*}

In order to identify the $\tau$ of the COVID-19 shock and subsequent recovery of the quality stocks, we have first identified the dominant IMF that fits the event as follows. We have calculated the correlation coefficient ($\nu$) between the original stock price and its IMFs and model-simulated stock price and its IMFs. We have also calculated variance ($\sigma^2$) of the IMFs as shown in Table~\ref{tab:1}. The value of $\nu$ measures the relationship between the individual IMF and the stock price. Whereas, the value of $\sigma^2$ measures the volatility of each IMF. From the Table~\ref{tab:1}, the values of $\nu$ and $\sigma^2$ shows that the IMF4 is the dominant mode for the original pharma and FMCG index and their model simulated data. All the four IMF4 modes along with their time series are shown in Fig~\ref{fig:tscaleV} (a), (b) (c) and (d), respectively. The average $\tau$ of the IMF4 for the pharma, FMCG index and simulated pharma and FMCG are approximately 57~$D$, 56~$D$, 58~$D$ and 58~$D$, respectively. Vertical lines in the figure show the bottom formation due to the COVID-19 shock. All the IMF4 shows that $\mathcal{T_S}$ and $\mathcal{T_R}$ are almost equal. For the pharma and FMCG stocks $\mathcal{T_S}\approx\mathcal{T_R}\approx 14~D.$ We have obtained that the $\mathcal{T_S}\approx\mathcal{T_R}$ for all the quality stocks that show V-shape recovery. It is pertinent to mention that there is no dominant mode present in the case of L-shape recovery.

\section{conclusion}
\label{sec:con}
In this paper, we have developed a model of the stock price movement during the COVID-19 shock and its subsequent recovery. The simulation is carried out assuming that during shock, the price crashes due to the fund outflow from the market, and does not depends on the financial antifragility ($\phi$) of a company. Whereas, the recovery of the stock price depends on the fund inflow towards a particular company or sector depending on $\phi$. The model simulates the stock price for different $\mathcal{T_S}$ and different $\phi$ using synthetic normalized net fund flow. The model reproduces the stock price movement during the pandemic very well. We have also identified the $\mathcal{T_S}$ and $\mathcal{T_R}$ from the model and original data using EMD based Hilbert Huang Transformation.

We obtained V-shape recovery of the quality stocks with positive $\phi$ using synthetic normalized net fund flow ($\Psi_{st}$).  The stock price recovers very well to its pre-shock value for the $\mathcal{T_S}=~20~D$ and 40~$D$ with fixed $\phi=0.4$. However, when the $\mathcal{T_S}$ extends beyond 60~D, it becomes difficult for the stock price to recover to its pre-shock price. We also obtained the V-shape recovery for $\phi=0.3,0.4, 0.5$ and 0.6 with fixed $\mathcal{T_S}=20~D$. The stock with a higher $\phi$ outperforms its peer with a lower $\phi$ after crises. Such performance in the stock price of certain quality stocks have been observed during the COVID-19 pandemic. In the case of the financially stressed stocks, i.e., with negative $\phi$, we obtained L-shape recovery of the stock price, and such stocks show higher negative depth in stock price with an increase in $\mathcal{T_S}.$ As the value of $\phi$ decreases, the duration of $\mathcal{T_S}$ increases that have been observed in several stocks during the pandemic.

We obtained V-shape recovery from the model using the normalized net fund flow $\Psi_t.$ In this simulation, we have used the average value of $\phi$ of the Pharma and FMCG. The simulated results are consistent with the original stock price movement of the Pharma and FMCG indices during the COVID-19 shock. On the other hand, for the companies with negative $\phi$, the model also consistent with L-shape movement of price. 

Finally, we obtained that the $\mathcal{T_S}$ and $\mathcal{T_R}$ of a quality stock during the COVID-19 is approximately equal. On the other hand, the companies with $\phi<0$ is yet to recover. The value of $\mathcal{T_S}$ and $\mathcal{T_R}$ for different sectors will be useful for making an investment decision. We observed that for some sectors like the banking where $\phi>0$, it still shows L-shape recovery. Such recovery depends on various other factors which will be studied in future work.

\section*{Acknowledgments}
The authors acknowledge Jean-Philippe Bouchaud, Dhurv Sharma and Paresh K. Narayan for their valuable comments and suggestions to improve the manuscript. NIT Sikkim is appreciated for allocating doctoral research fellowships to A.M. and A.R.

\bibliography{apssamp}

\end{document}